\documentclass[preprint,noshowpacs,preprintnumbers]{revtex4}
\usepackage{graphicx}
\usepackage{dcolumn}
\usepackage{bm}

\begin{document}

\title{Multiplicity and transverse energy of produced gluon in relativistic
heavy ion collision}
\author{Bo-Wen Xiao}
\email{bowen@phys.columbia.edu}
\affiliation{Department of Physics, Columbia University, New York, NY, 10027, USA}

\begin{abstract}
We present a simple gluon production picture which is based on the
McLerran-Venugopalan model and gluon BFKL evolution in relativistic heavy
ion collision. Results for the multiplicity and transverse energy
distribution in both the central and forward rapidity regions for gluon
production in early stages of heavy ion collisions at the LHC are given.
Finally, we provide a general qualitative discussion of the consequences of
the forward rapidity behavior of produced gluons.
\end{abstract}

\date{\today}
\maketitle

\vfill

\vfill


\section{Introduction}

The description of the early stages of heavy ion collisions before
hadronization is currently an interesting and evolving topic in heavy ion
physics, and it is still not clear how good a description can be obtained
from Quantum Chromodynamics(QCD). However, a better understanding of the
properties of the initially produced gluons would shed some light on the
early stages of the physics in heavy ion collisions.

The McLerran-Venugopalan model\cite{MV94} describes nuclear collisions at
the Relativistic Relativistic Heavy Ion Collider (RHIC) in terms of the
gluons in semi-classical non-abelian Weizs\"{a}cker-Williams field generated
by the valence quarks of relativistic nuclei. At RHIC energies, it appears
to be a good approximation since the actual gluon occupation number(gluon
phase space density) is large enough to be described by the classical
non-abelian Weizs\"{a}cker-Williams field. Specifically, we write the
occupation number as $f_{g}=\frac{(2\pi )^{3}}{2(N_{c}^{2}-1)}\frac{dN_{g}}{%
d^{3}bd^{3}k}=\frac{(2\pi )^{3}}{2(N_{c}^{2}-1)}\frac{dNg}{%
dyd^{2}bd^{2}k_{\perp }}$. At LHC energies, quantum evolution in the
longitudinal momentum (BFKL evolution\cite{BFKL}) should be more important
and increase the number of gluons by increasing the saturation momentum
while the occupation number $f_{g}$ does not change much\cite{Mue2002} (we
can see this either from the McLerran-Venugopalan model or from the
discussion which follows.)

The Balisky, Fadin, Kuraev and Lipatov (BFKL)\cite{BFKL} equation is an
equation which describes QCD hard scattering processes with a single hard
transverse momentum scale and sums all terms like $(\alpha \ln(1/x))^n$(or $%
(\alpha Y)^n$). Explicit evaluation of this equation tells us that how
nucleus evolves when it is boosted to a larger rapidity, and this gives the
rapidity evolution of the nuclear gluon distribution. Moreover, it has been
shown\cite{Mue03} that the McLerran-Venugopalan model and fixed coupling
BFKL evolution in the vicinity of the saturation boundary predicts
geometrical scaling as shown below in Eq.~(\ref{Geometry}). We suppose that
this geometrical scaling is the right description for the nuclear gluon
distribution beyond the saturation momentum, that is when $k_{\perp}>Q_{s}$,
at LHC energies.

Based on a parton saturation model\cite{Gri83,Blai87,Mue86} and the color
glass condensate\cite{MV94,Cgc,Cgc2}, as well as a parton-hadron duality
hypothesis, Kharzeev $\mathit{et al}$ \cite{Khar} have successfully made
consistent calculation of hadron multiplicity in heavy ion collision, by
assuming the produced gluons in the initial state hadronize into a
corresponding number of hadrons in the final state. It appears that
saturation and the Color Glass Condensate provide a reasonable physical
pictures for relativistic heavy ion collisions. Therefore, these tools have
begun to be used to describe and understand the thermalization stages of the
produced gluons right after their formation during the heavy ion collision.
Thus, a description of the produced gluon multiplicity and transverse energy
distribution throughout all rapidity regions seems to be more and more
important, and this is part of the motivation of this work.

In this paper, we are interested in the initial production processes of
gluons in a heavy ion collision, which come from the freeing of the virtual
gluons in colliding nuclei. The essential feature of this approach is that
we choose our reference frame so that the produced gluon on which we are
focusing has the rapidity $y$. Thus each of these two colliding nuclei have
BFKL evolution in this particular frame. Based on the BFKL evolved gluon
distribution in the nuclei and the approximate criteria\cite{Blai87,Mue00}
developed by Blaizot and Mueller, we calculate the distribution of freed
gluons from four different possible regions which are shown in Eqs.~(\ref{N1}%
),(\ref{N2}),(\ref{N3}) and (\ref{N4}). In the end, we sum these
contribution and obtain the total multiplicity and transverse energy
distribution in Eqs.~(\ref{N}) and (\ref{E}) for all produced gluons. In
Eq.~(\ref{N}), we find that the first and the third terms are similar to the
results in ref.\cite{Khar} while the second term is different since we use
geometrical scaling in the region beyond the saturation momentum. We also
calculate the average transverse momentum per gluon in forward rapidity
region. The calculations show that the average transverse momentum increases
a lot while the number of produced gluons decreases rapidly in the forward
rapidity region which may lead to some interesting physics in that region.
As we will discuss in detail later, we consider the consequence of this
result for jet quenching, gluon thermalization and elliptic flow. One of the
conclusions is that we might expect the jet quenching effect to be less
significant in forward rapidity region.

In section 2, we give a brief review of the McLerran-Venugopalan model, and
derive the gluon distribution function in nuclei by following the
interpretation of this model provided in ref.\cite{Mue98}.

In section 3, we boost both of these two colliding nuclei (the projectile
and the target) away from central rapidity region, and consider BFKL
evolution\cite{BFKL} for both nuclei.

In section 4, based on the framework established in above sections and the
approximate criteria\cite{Blai87,Mue00} for gluon freeing, we calculate the
rapidity dependence of the multiplicity and of the transverse energy of
freed gluons.

Finally, in section 5, some qualitative discussions regarding the
consequences of the forward rapidity behavior of produced gluons as well as
a brief summary are provided.

\section{McLerran-Venugopalan model}

Mueller and Kovechegov have shown\cite{Mue98} that in the
McLerran-Venugopalan model, the gluon distribution of the relativistic
nucleus can be written as following:

\begin{equation}
\frac{dxG(x,k_{\perp }^{2})}{d^{2}bd^{2}k_{\perp }}=\frac{N_{c}^{2}-1}{4\pi
^{4}\alpha N_{c}}\int \frac{d^{2}x_{\perp }}{\underline{x_{\perp}}^{2}}%
\left( 1-e^{-\underline{x}^{2}Q_{s}^{2}/4}\right) e^{-i\underline{k_{\perp}}%
\cdot \underline{x_{\perp }}},  \label{G}
\end{equation}%
in which the $k_{\perp }$ is the transverse momentum of the gluon, and the $%
x=\frac{k^{+}}{P^{+}}$ is the longitudinal momentum fraction of the gluon.
The saturation momentum, $Q_{s}$, is given by
\begin{equation}
Q_{s}^{2}=\frac{8\pi ^{2}\alpha N_{c}}{N_{c}^{2}-1}\sqrt{R^{2}-b^{2}}\rho
xg(x,4/\underline{x}^{2}),  \label{Q}
\end{equation}%
in which the $xg(x,4/\underline{x}^{2})$ is just the gluon distribution of a
single nucleon inside this nucleus. In the region $k_{\perp }^{2}\ll
Q_{s}^{2}$, one can neglect the $\underline{x}^{2}$ dependence of the $%
Q_{s}^{2}$ and evaluate the above gluon distribution integral
\begin{equation}
\frac{dxG(x,k_{\perp }^{2})}{d^{2}bd^{2}k_{\perp }}=\frac{N_{c}^{2}-1}{4\pi
^{3}\alpha N_{c}}\int_{1}^{\infty }\frac{dt}{t}e^{-tk_{\perp
}^{2}/Q_{s}^{2}}=\frac{N_{c}^{2}-1}{4\pi ^{3}\alpha N_{c}}\ln \left( \frac{%
Q_{s}^{2}}{k_{\perp }^{2}}\right).  \label{G1}
\end{equation}%
Moreover, for a large nucleus using BFKL evolution, it has been found\cite%
{Mue99,Mue02A,Ian01} that:
\begin{equation}
\frac{dxG(x,k_{\perp }^{2})}{d^{2}bd^{2}k_{\perp }}=\frac{N_{c}^{2}-1}{4\pi
^{3}\alpha N_{c}}\frac{1-\lambda _{0}}{2\chi (\lambda _{0})}\ln \left( \frac{%
Q_{s}^{2}}{k_{\perp }^{2}}\right) ,  \label{G2}
\end{equation}%
where $\lambda _{0}$ is defined by $\chi ^{\prime }(\lambda _{0})=-\frac{%
\chi (\lambda _{0})}{1-\lambda _{0}}$, and $\chi (\lambda _{0})=\psi (1)-%
\frac{1}{2}\psi (\lambda _{0})-\frac{1}{2}\psi (1-\lambda _{0})$ with the $%
\psi =\frac{d\ln \Gamma (\lambda )}{d\lambda }$ defined as the logarithmic
derivative of the gamma-function. It is likely that the factors $ \frac{%
N_{c}^{2}-1}{4\pi ^{3}\alpha N_{c}}$ and $\ln \left( \frac{Q_{s}^{2}}{%
k_{\perp }^{2}}\right) $ in the gluon distribution formula can be taken as
general results except for an overall factor which could be fixed by a
comparison to RHIC data. The essential physical picture is that the gluon
density per unit area and per unit of two-dimension transverse momentum,
that is per unit of transverse phase space, is limited by (or saturates at)
the product of a constant and logarithmic factor $\ln \left( \frac{Q_{s}^{2}%
}{k_{\perp }^{2}}\right) $ with a upper momentum limit $Q_{s}^{2}$.
Therefore, one can write the gluon phase space density as
\begin{equation}
\frac{dxG(x,k_{\perp }^{2})}{d^{2}bd^{2}k_{\perp }}=\frac{k}{c}\frac{%
N_{c}^{2}-1}{4\pi ^{3}\alpha N_{c}}\ln \left( \frac{Q_{s}^{2}}{k_{\perp }^{2}%
}\right) ,  \label{G3}
\end{equation}%
in which $k$ is a $O(1)$ constant, and $c=\frac{2\chi (\lambda _{0})}{%
1-\lambda _{0}}=4.88$. In practice, one can change above gluon distribution
into a uniform distribution as $\frac{dxG(x,k_{\perp }^{2})}{%
d^{2}bd^{2}k_{\perp }}=\frac{k}{c}\frac{N_{c}^{2}-1}{4\pi ^{3}\alpha N_{c}}$%
, since the logarithmic factor disappears after integration over $k_{\perp}$.

\section{BFKL evolution and geometrical scaling}

However, the simple McLerran-Venugopalan (MV) gluon distribution does not
contain any evolution in Bjorken $x$ which is necessary to explore the
transverse energy spectrum. In order to extend our discussion from the
central rapidity region to the forward rapidity region, we need to consider
BFKL evolution of the saturation momentum. The leading order BFKL evolution
for the saturation momentum is
\begin{equation}
Q^{2}_{s}(A,Y)=Q^{2}(A)\frac{\exp \left[ \frac{2\alpha N_{c}}{\pi }\frac{%
\chi (\lambda _{0})}{1-\lambda _{0}}Y\right] }{\left( \alpha Y\right) ^{%
\frac{3}{2(1-\lambda _{0})}}}.  \label{S1}
\end{equation}%
Numerical calculation shows that $\lambda _{0}=0.372$ and $c=\frac{2\chi
(\lambda _{0})}{1-\lambda _{0}}=4.88$. This leads to the saturation momentum
$Q_{s}^{2}(Y)$ goes like $\exp \left[ \lambda _{s}Y\right] $ with $\lambda
_{s}\sim 1$ if taking $\alpha =1/3$. This value is much larger than the
phenomenological approaches\cite{Hera} which find that $\lambda _{s}\ $is
about $0.3$. Fortunately, the next to leading order calculation \cite{Tri02}
shows that $\lambda _{s}$ effectively behaves like a constant around $0.30$
since the next to leading correction slows down the evolution. Therefore, we
shall take
\begin{equation}
Q_{s}(A,y)^{2}=Q(A)^{2}\exp \left[ 0.30y\right]  \label{S2}
\end{equation}
in our numerical calculations.

Nevertheless, the above simple MV gluon distributions Eqs.~(\ref{G1}),(\ref%
{G2}) and (\ref{G3}) are only valid when $k_{\perp}<<Q_{s}$, and they do not
tell us how it behaves when the gluon transverse momentum $k_{\perp }$ is
larger than the saturation momentum. Therefore, we need to consider the
small-x evolution\cite{BFKL,Mue03} of the saturation momentum $Q_{s}$ and
the gluon distribution in the vicinity of the saturation boundary. This
gives a geometrical scaling of the gluon distribution beyond the saturation
momentum scale \cite{Mue02,Mue03}. It has been shown \cite{Mue03} that
beyond the saturation momentum, when $k_{\perp } > Q_{s}$, the evolved gluon
distribution behaves like
\begin{equation}
\frac{dxG(x,k_{\perp }^{2})}{d^{2}bd^{2}k_{\perp }}=k\frac{N_{c}^{2}-1}{4\pi
^{3}\alpha N_{c}}\frac{1}{c}\left( \frac{Q_{s}^{2}}{k_{\perp }^{2}}\right)
^{1-\lambda _{0}}\ln \left( \frac{k_{\perp }^{2}}{Q_{s}^{2}}\right) .
\label{Geometry}
\end{equation}
We have chosen the same overall constant to agree with Eq.~(\ref{G3}).

\section{Produced gluon multiplicity and transverse energy}

In this section we use the McLerran-Venugopalan model together with BFKL
evolution to investigate behavior of produced gluons in all rapidity
regions. Let us concentrate on a simple case first, in which we consider the
head on collision between the projectile(left-moving) nucleus and
target(right-moving) nucleus where both of the nuclei have the same number
of nucleon; therefore the physics is symmetric under $y\Longleftrightarrow -y
$. We also do not consider the impact parameter or centrality dependence for
the moment. Thus for now we only consider the $y>0$ region.

\begin{figure}[tbp]
\par
\begin{center}
\includegraphics[width=9cm]{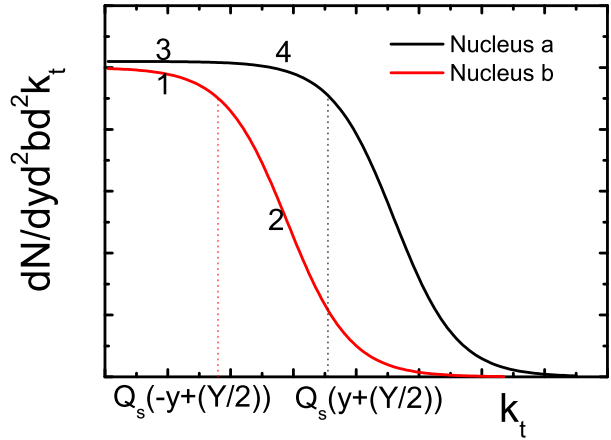}
\end{center}
\caption[*]{\baselineskip13pt Gluon distribution for two BFKL evolved
nuclei, in which we denote the projectile(left-moving) nucleus saturation
momentum as $Q_{s}(y+\frac{Y}{2})$ and target(right-moving) nucleus
saturation momentum as $Q_{s}(-y+\frac{Y}{2})$. }
\label{gluons}
\end{figure}

\begin{figure}[tbp]
\par
\begin{center}
\includegraphics[width=12cm]{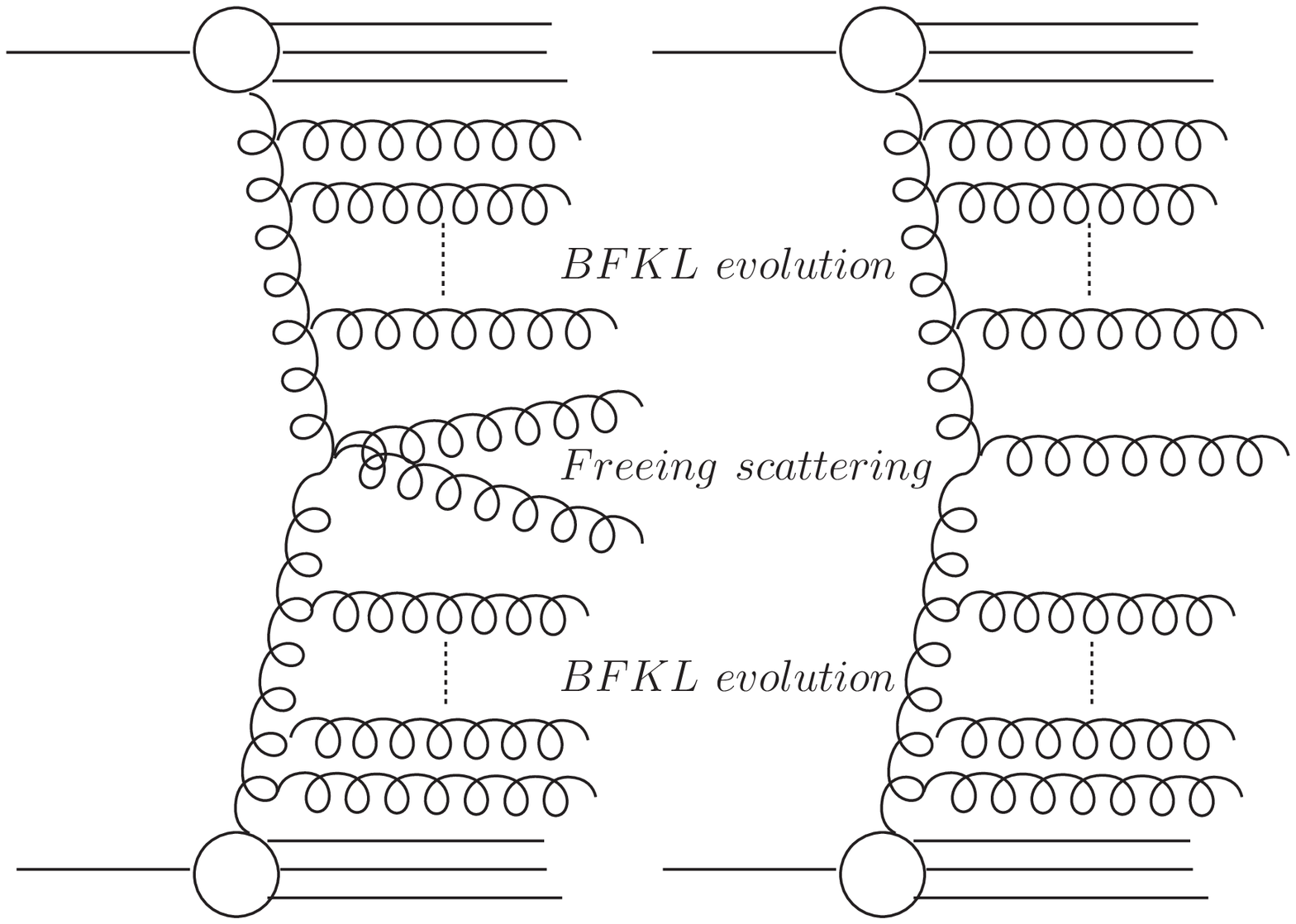}
\end{center}
\caption[*]{\baselineskip13pt Gluon BFKL evolution in the projectile and
target nuclei as well as two possible freeing scattering process during the
collisions. }
\label{freeing}
\end{figure}

The essential feature of this approach is that we choose a reference frame
in which the produced gluons has zero rapidity. The two colliding nuclei
then are assumed to have saturation momenta in the BFKL evolution region.
Suppose the total rapidity difference between the target and projectile is $Y
$, then the saturation momenta for the target and projectile nuclei are $%
Q_{s}(-y+\frac{Y}{2})$ and $Q_{s}(y+\frac{Y}{2})$ in this reference frame,
respectively. Based on the BFKL evolved gluon distribution in the nuclei, we
can determine the distribution of the freed gluons in our reference frame by
using an approximate criteria\cite{Blai87,Mue00} developed by Blaizot and
Mueller just as one usually does in the central rapidity region in the
center of mass frame. The approximate criteria states that the gluons which
receive enough transverse momentum to push them onto mass shell will be
freed during the collision. As far as we know, the typical momentum of the
gluons inside the nuclei are $Q_{s}(-y+\frac{Y}{2})$ and $Q_{s}(y+\frac{Y}{2}%
)$ respectively. Correspondingly, the necessary typical transfer momenta in
order to free gluons are $Q_{s}(-y+\frac{Y}{2})$ and $Q_{s}(y+\frac{Y}{2})$
respectively for these two nuclei as well. In this sense, we can separate
gluons inside the projectile and the target into four different according to
these two saturation momenta as shown in Fig.(\ref{gluons}).

1. Gluons inside the target with the typical saturation momentum $Q_{s}(-y+%
\frac{Y}{2})$.

a. Gluons inside the target with a transverse momentum $k_{t}<Q_{s}(-y+\frac{%
Y}{2})$ (Noting that $Q_{s}(-y+\frac{Y}{2})<Q_{s}(y+\frac{Y}{2})$ when $y>0$%
) are labeled as 1 in Fig.(\ref{gluons}). They scatter with the gluons in
the projectile nucleus which have a typical momentum $Q_{s}(y+\frac{Y}{2})$
due to the gluon saturation and BFKL evolution which are illustrated in Fig.(%
\ref{freeing}). During the scattering, the virtual gluons which are
originally bounded inside the target nucleus are freed according to the
approximate criteria that we have stated above. Therefore, those gluons in
part $1$ inside the target can be freed (or more properly, they are
produced), and we calculate their total number per unit rapidity per unit
area from the definition $\frac{dN_{1}}{dyd^{2}b}=\int%
\nolimits_{0}^{Q_{s}(-y+\frac{Y}{2})}d^{2}k_{\perp }\frac{dxG(x,k_{\perp
}^{2})}{d^{2}bd^{2}k_{\perp }}$ by integrating Eq.~(\ref{G2} )
\begin{equation}
\frac{dN_{1}}{dyd^{2}b}=\frac{k}{c}\frac{N_{c}^{2}-1}{4\pi ^{3}\alpha N_{c}}%
\pi \left( Q_{s}(-y+\frac{Y}{2})\right) ^{2}.  \label{N0}
\end{equation}
Then, how are these produced gluons distributed in momentum space?
Originally, we assume these gluons are almost uniformly distributed from QCD
energy scale to the saturation momentum $Q_{s}(-y+\frac{Y}{2})$ according to
Eq.~(\ref{G2}) and its simplified form where in Eq.~(\ref{G2}) the
logarithmic term is replaced by a constant. During the freeing process,
these gluons exchange transverse momentum in the scattering with gluons from
the projectile which have a typical momentum $Q_{s}(y+\frac{Y}{2})$. This
means that the momentum space distribution of these gluons has been enlarged
from $Q_{s}(-y+\frac{Y}{2})$ to $Q_{s}(y+\frac{Y}{2})$. Therefore, we
presume that these gluons would be uniformly distributed in the momentum
space from very small transverse momentum (could be close to zero) to the
typical momentum of the projectile nucleus $Q_{s}(y+\frac{Y}{2})$ scale
which is the typical momentum the freed gluons can acquire after the
freeing. Then one finally obtain:%
\begin{equation}
\frac{dN_{1}}{dyd^{2}bd^{2}k_{\perp }}=\frac{k}{c}\frac{N_{c}^{2}-1}{4\pi
^{3}\alpha N_{c}}\left( \frac{Q_{s}(-y+\frac{Y}{2})}{Q_{s}(y+\frac{Y}{2})}%
\right) ^{2},  \label{N1}
\end{equation}%
with $k_{t}<Q_{s}(y+\frac{Y}{2})$. Note that the distribution of the freed
gluons has been broadened from $k_{t}<Q_{s}(-y+\frac{Y}{2})$ to $%
k_{t}<Q_{s}(y+\frac{Y}{2})$ due to the scattering with the gluons from the
projectile.

b. Gluons inside the target with a transverse momentum $k_{t}$ obeying $%
Q_{s}(-y+ \frac{Y}{2})<k_{t}<Q_{s}(y+\frac{Y}{2})$ are labeled as 2 in Fig.(%
\ref{gluons}). They scatter with the gluons in the projectile nucleus which
have a typical momentum $Q_{s}(y+\frac{Y}{2})$. In this region, we believe
that the distribution of the gluons inside the target obeys the geometrical
scaling which we have shown in Eq.~(\ref{Geometry}). Therefore, these gluons
can be freed and we assume they are uniformly distributed from very small
transverse momentum to the typical momentum scale of the projectile nucleus $%
Q_{s}(y+\frac{Y}{2})$ scale after the freeing. The case is almost the same
as the case a above except for the initial distribution of the gluons in the
target. Then, we integrate over this region, and get the freed gluon
distribution
\begin{equation}
\frac{dN_{2}}{dyd^{2}bd^{2}k_{\perp }}=\frac{k}{c}\frac{N_{c}^{2}-1}{4\pi
^{3}\alpha N_{c}}\left\{ \frac{1}{\lambda _{0}}\left( \left( \frac{Q_{s}(y+%
\frac{Y}{2})}{Q_{s}(-y+\frac{Y}{2})}\right) ^{2\lambda _{0}}-1\right)
\right\} \left( \frac{Q_{s}(-y+\frac{Y}{2})}{Q_{s}(y+\frac{Y}{2})}\right)
^{2}.  \label{N2}
\end{equation}%
with $k_{t}<Q_{s}(y+\frac{Y}{2})$. Actually, we neglect the logarithmic term
again in Eq.~(\ref{Geometry}) before arriving at the result in $\left\{
{}\right\} $ in above equation while the exact result would be $\frac{1}{%
\lambda _{0}^{2}}\left[ \left( \frac{Q_{s}(y+\frac{Y}{2})}{Q_{s}(-y+\frac{Y}{%
2})}\right) ^{2\lambda _{0}}\left( \ln \left( \frac{Q_{s}(y+\frac{Y}{2})}{%
Q_{s}(-y+\frac{Y}{2})}\right) ^{2\lambda _{0}}-1\right) +1\right] $. This
simplification does not affect the physics much. In addition, we want to
comment on the result that this is the largest contribution in the
moderately large rapidity region due to the $\left( \frac{Q_{s}(y+\frac{Y}{2}%
)}{Q_{s}(-y+\frac{Y}{2})}\right) ^{\lambda _{0}}$ term which differs from
the result in \cite{Khar} although we both agree that this region gives
largest contribution. This difference comes from the geometrical scaling,
and we expect that heavy ion collisions at LHC might observe this scaling
while at RHIC geometrical scaling may not yet set in since the saturation
momentum may not be large enough.

c.Gluons inside the target with $k_{t}>Q_{s}(y+\frac{Y}{2})$. In this
region, the gluons are very few and the parton densities in the target are
in the DGLAP evolution region. We believe this region does not contribute
much.

2. Gluons inside the projectile with the typical saturation momentum $%
Q_{s}(y+\frac{Y}{2})$.

a. Gluons inside the projectile with a transverse momentum $k_{t}<Q_{s}(-y+%
\frac{Y}{2})$, labeled as 3 in Fig.(\ref{gluons}), scatter with the gluons
in the target nucleus which have a typical momentum $Q_{s}(-y+\frac{Y}{2})$.
Therefore, these gluons can be freed and we assume they uniformly
distributed from very small transverse momentum to the typical momentum
scale of the target nucleus $Q_{s}(-y+\frac{Y}{2}) $ after the freeing. Then
one finds
\begin{equation}
\frac{dN_{3}}{dyd^{2}bd^{2}k_{\perp }}=\frac{k}{c}\frac{N_{c}^{2}-1}{4\pi
^{3}\alpha N_{c}},  \label{N3}
\end{equation}
with $k_{t}<Q_{s}(-y+\frac{Y}{2})$. Note that the range of the momentum
remains the same because the typical momentum that the freed gluons scatter
with is $Q_{s}(-y+\frac{Y}{2})$. If we integrate over two-dimension phase
space in Eqs.~(\ref{N1}) and (\ref{N3}), we get the same results with
Kharzeev $\mathit{etal}$ \cite{Khar}.

b.Gluons inside the projectile with a transverse momentum $Q_{s}(-y+\frac{Y}{%
2})<k_{t}<Q_{s}(y+\frac{Y}{2})$ are labeled as 4 in Fig.(\ref{gluons}). They
scatter with the gluons in the target nucleus whose distribution is given by
geometrical scaling. The problem here is very similar to that of case 1.b.
and we presume that the contribution for this part can be written as
\begin{equation}
\frac{dN_{4}}{dyd^{2}bd^{2}k_{\perp }}=\delta \frac{dN_{2}}{%
dyd^{2}bd^{2}k_{\perp }} .  \label{N4}
\end{equation}%
with $0<\delta <1$ and $k_{t}<Q_{s}(y+\frac{Y}{2})$.

c.Gluons inside the projectile with $k_{t}>Q_{s}(y+\frac{Y}{2})$. In this
region the gluons are very few and the parton densities in the projectile
are in the DGLAP evolution region. We believe that this region does not
contribute much.

Now, when we generalize our discussion to the very forward rapidity region,
the factor $x=\frac{Q_{s}(A,\pm y+\frac{Y}{2})}{\sqrt{s}}\exp \left[ y\right]
$ is no longer small. In this case we need to include a limiting
fragmentation factor $(1-x)^{4}$ which comes from the phenomenological quark
counting rules\cite{Bro73}.

Then, we add all contribution together and integrate over the two-dimension
phase space to get the total multiplicity with the fragmentation factor
taken into account as a large $x$ correction
\begin{eqnarray}
\frac{dN}{dy} &=&\sum\limits_{i=1}^{4}\int d^{2}bd^{2}k_{\perp }\frac{dN_{i}%
}{dyd^{2}bd^{2}k_{\perp }}(1-x)^{4} \\
&=&\left( \frac{k}{c}\frac{N_{c}^{2}-1}{4\pi ^{3}\alpha N_{c}}\right) \left(
\pi R^{2}\right) \left( \pi Q_{s}^{2}(A,-y+\frac{Y}{2})\right)  \nonumber \\
&&\times \left\{ \left( 1-\frac{Q_{s}(-y+\frac{Y}{2})}{\sqrt{s}}e^{y}\right)
^{4}\right.  \nonumber \\
&&\left. +\left[ \frac{1+\delta }{\lambda _{0}}\left( \left( \frac{Q_{s}(y+%
\frac{Y}{2})}{Q_{s}(-y+\frac{Y}{2})}\right) ^{2\lambda _{0}}-1\right) +1%
\right] \left( 1-\frac{Q_{s}(y+\frac{Y}{2})}{\sqrt{s}}e^{y}\right)
^{4}\right\} .  \label{N}
\end{eqnarray}
In the above equation we have made the approximation that $x=\frac{k_{\perp }%
}{\sqrt{s}}\exp \left[ y\right]$ for the gluons with transverse momentum $%
k_{\perp }$ in the integral is so small that $(1-x)^{4}$ is always around $1$
until $k_{\perp }$ approaches the upper limits $Q_{s}(\pm y+\frac{Y}{2})$.

Beginning with the approximate criteria of freeing gluon, we finally arrive
at almost the same result as Kharzeev $\mathit{etal}$ \cite{Khar} except for
the second term in above equation since we are using the geometrical scaling
gluon distribution when $k_{\perp }>Q_{s}$.(We believe that the geometrical
scaling distribution might be the right picture to describe gluon
distribution beyond the saturation momentum at LHC energies\cite{Mueller}.)
Moreover, we also notice that the second term, which is the contribution
from Eq.~(\ref{N2}), is the largest term in forward rapidity region since it
is enhanced by a factor of $\left( \frac{Q_{s}(y+\frac{Y}{2})}{Q_{s}(-y+%
\frac{Y}{2})}\right) ^{2\lambda _{0}}$. Then we can go one more step
further, averaging over different transverse momenta to get the transverse
energy distribution,
\begin{eqnarray}
\frac{dE_{T}}{dy} &=&\sum\limits_{i=1}^{4}\int d^{2}bd^{2}k_{\perp }\frac{%
dN_{i}}{dyd^{2}bd^{2}k_{\perp }}(1-x)^{4}k_{\perp } \\
&=&\left( \frac{k}{c}\frac{N_{c}^{2}-1}{4\pi ^{3}\alpha N_{c}}\right) \left(
\pi R^{2}\right) \frac{2}{3}\left( \pi Q_{s}(A,-y+\frac{Y}{2})^{2}\right)
\nonumber \\
&&\times \left\{ Q_{s}(-y+\frac{Y}{2})\left( 1-\frac{Q_{s}(-y+\frac{Y}{2})}{%
\sqrt{s}}e^{y}\right) ^{4}\right.  \nonumber \\
&&\left. +Q_{s}(y+\frac{Y}{2})\left[ \frac{1+\delta }{\lambda _{0}}\left(
\left( \frac{Q_{s}(y+\frac{Y}{2})}{Q_{s}(-y+\frac{Y}{2})}\right) ^{2\lambda
_{0}}-1\right) +1\right] \left( 1-\frac{Q_{s}(y+\frac{Y}{2})}{\sqrt{s}}%
e^{y}\right) ^{4}\right\} .  \label{E}
\end{eqnarray}

\begin{figure}[tbp]
\par
\begin{center}
\includegraphics[width=10cm]{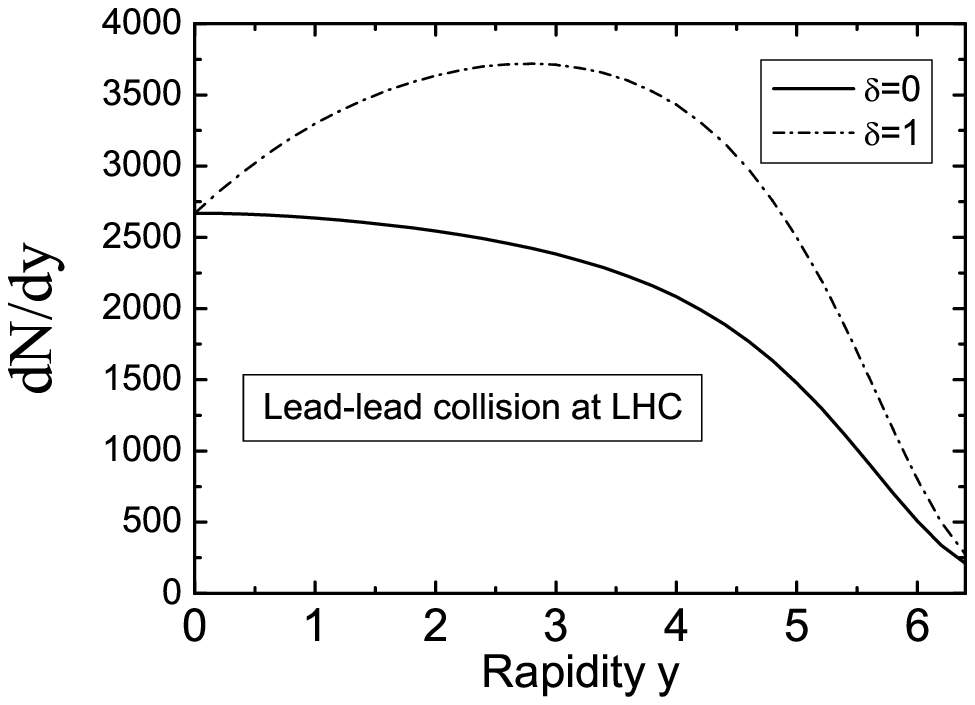} \includegraphics[width=10cm]{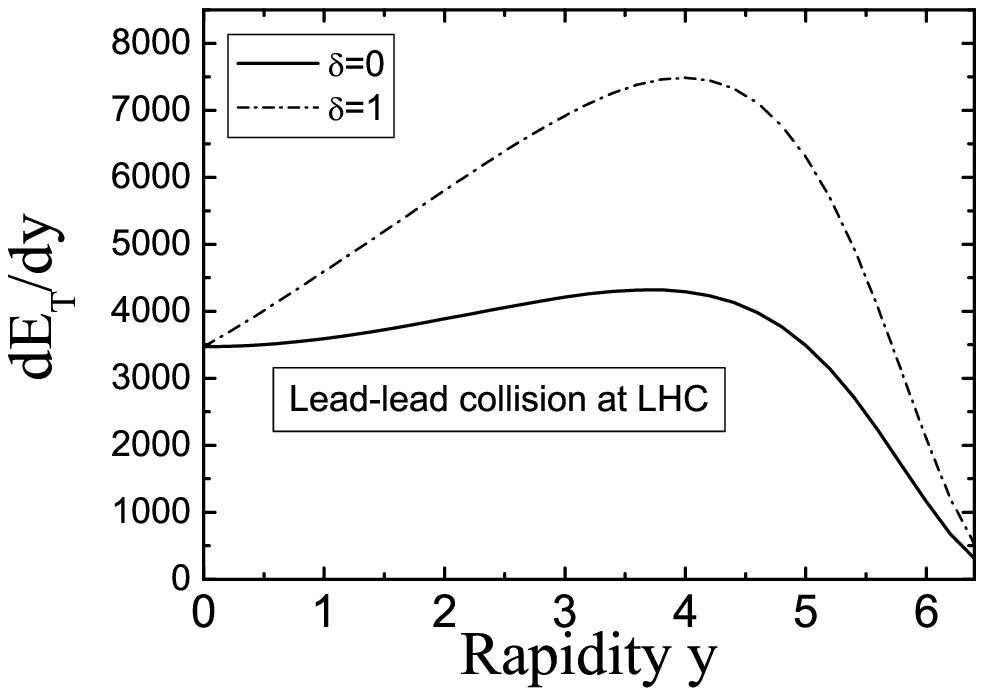}
\includegraphics[width=10cm]{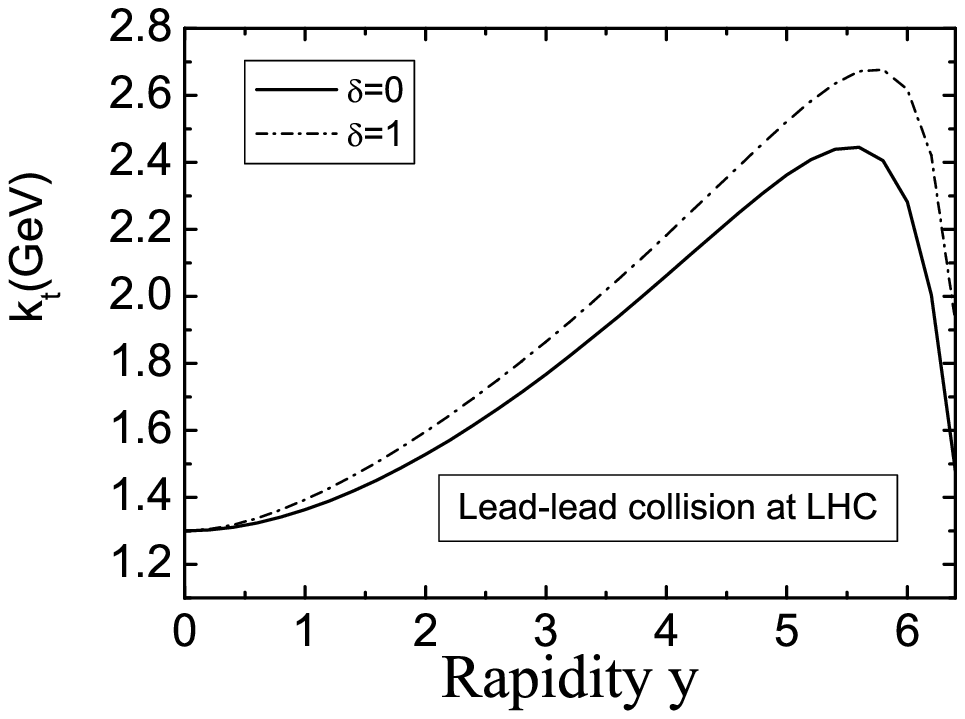}
\end{center}
\caption[*]{\baselineskip13pt Predictions of the multiplicity $\frac{dN}{dy}$%
, transverse energy distribution $\frac{dE_{T}}{dy}$ and average transverse
momentum per gluon $\langle k_{\perp}(y) \rangle$ for the $\protect\sqrt{s}%
=5500GeV$ lead-lead head-on collision at LHC when using $Q_s =1.95GeV$ and $%
\protect\alpha =1/3$, respectively.}
\label{dy}
\end{figure}

In addition, we should fix the parameters which appear in the multiplicity
and transverse energy distribution before we plot $\frac{dN}{dy}$ and $\frac{%
dE_{T}}{dy}$ as a function of rapidity $y$ in fig.~(\ref{dy}). We take the
parameters to be $\alpha =1/3$ and $N_{c}=3$ according to convention, and we
allow the parameter $\delta $ to vary from $0$ to $1$ in the plots.

Noting that $\exp [\frac{\lambda _{s}\,Y}{2}]=\left( \frac{\sqrt{s}}{M_{nuc}}%
\right) ^{\lambda _{s}}$ where $M_{nuc}$ is the rest mass of one of the
colliding nuclei, we can write the Eq.(\ref{S2}) as
\begin{equation}
Q_{s}^{2}(A,y+\frac{Y}{2})=Q_{s_{0}}^{2}(A)\left( \frac{\sqrt{s}}{\sqrt{s_{0}%
}}\right) ^{\lambda _{s}}\exp \left[ \lambda _{s}y\right] .  \label{s3}
\end{equation}%
Then, phenomenologically we take:
\begin{eqnarray}
R &=&r_{0}A^{\frac{1}{3}},  \label{R} \\
Q_{s}^{2}(A) &=&(1.1GeV)^{2}\left( \frac{A}{197}\right) ^{\frac{1}{3}}\left(
\frac{\sqrt{s}}{130GeV}\right) ^{\lambda _{s}}  \label{QS}
\end{eqnarray}%
where $r_{0}=1.2fm$ is obtained from the empirical nuclear radius formula
and $Q_{s}\left\vert _{A=197}\right. =1.1GeV$ \cite{Mue2002} is an
estimation based on RHIC Au-Au collision at $\sqrt{s_{0}}=130GeV$ by using
Eq.~(\ref{Q}) and taking $\rho =\frac{A}{\frac{4}{3}\pi R^{3}}=\frac{1}{%
\frac{4}{3}\pi r_{0}^{3}}$, $xg(x,Q^{2})=3.3$. Therefore, we predict that
the saturation momentum at LHC is around $1.95GeV$ at $\sqrt{s}=5500GeV$ in
the central rapidity region and $\pi R^{2}=158fm^{2}$ for lead nuclei from
Eqs.~(\ref{R}) and (\ref{QS}). Noting that the geometrical scaling
contribution in the multiplicity Eq.~(\ref{N}) vanishes in the $y=0$ central
rapidity region means that $\frac{dN}{dy}\left\vert _{y=0}\right. $\ should
match RHIC data. Thus, we can fix the ratio of parameters $k$ and $c$ by
using the multiplicity data in the central rapidity region from PHOBOS \cite%
{Back:2001bq} which provides $\frac{dN_{ch}}{dy}\left\vert _{y=0}^{\sqrt{s}%
=130GeV}\right. =547\pm 55$ or $\frac{dN}{dy}\left\vert _{y=0}^{\sqrt{s}%
=130GeV}\right. =821\pm 83$ (Here we assume $\frac{dN}{dy}\left\vert
_{y=0}\right. =\frac{3}{2}\frac{dN_{ch}}{dy}\left\vert _{y=0}\right. $in
which $\frac{dN_{ch}}{dy}\left\vert _{y=0}\right. $ is the multiplicity of
charged particles, and we also neglect the difference between the rapidity $y
$ and pseudorapidity $\eta $.). This helps us to fix $\frac{k}{c}=\frac{1}{%
2.34}$ which agrees with Mueller's conjecture\cite{Mue00}(Mueller
suggests that the total number of the produced gluons is
proportional to the total number of the gluons in the initial
nuclear wave-function and the proportionality coefficient $\epsilon$
is called gluon liberation coefficient.) and previous
studies\cite{Krasnitz:1998ns,Kovchegov:2000hz,Krasnitz:2002mn,Lappi:2003bi}
that the gluon liberation coefficient $\epsilon
=\frac{2k}{c}=\frac{2}{2.34}$ (this relation can be
extracted from Eq.~(\ref{N}) via the definition $\frac{dN}{d^{2}bdy}%
\left\vert _{y=0}\right. =\epsilon \frac{N_{c}^{2}-1}{4\pi ^{3}\alpha N_{c}}%
\pi Q_{s}^{2}$ at central rapidity region.) is an order $1$
coefficient.

Then with fixed coefficients and Eq.~(\ref{N}), we are able to predict that:
$\frac{dN_{ch}}{dy}\left\vert _{y=0}^{\sqrt{s}=200GeV}\right. =630$ which
agrees with PHOBOS data\cite{Back:2001ae} $\frac{dN_{ch}}{dy}\left\vert
_{y=0}^{\sqrt{s}=200GeV}\right. =650\pm 35(syst)$ and $\frac{dN_{ch}}{dy}%
\left\vert _{y=0}^{\sqrt{s}=5500GeV}\right. =1780$ for the lead-lead
collision in the LHC energies, which is close to Kharzeev's
prediction\cite{Khar}($\frac{dN_{ch}}{dy}%
\left\vert _{y=0}^{\sqrt{s}=5500GeV}\right. =1750-2100$).

Finally, for each rapidity, we calculate the average transverse momentum per
gluon as
\begin{equation}
\langle k_{\perp}(y) \rangle=\frac{\frac{dE_{T}}{dy}}{\frac{dN}{dy}}
\end{equation}
and plot $\langle k_{\perp }(y)\rangle $ also in Fig.~(\ref{dy}).

We find that the average transverse momentum per parton in our calculation
(as well as that in Kharzeev's paper if one generalize it and calculate the
average transverse momentum) increases a lot as we move to the forward
rapidity region.(When $y>6$, $\langle k_{\perp }(y)\rangle $ starts to
decrease because of the nonlinear effect of the limiting fragmentation
factor $(1-x)^{4}$. We do not expect our discussion to hold in this region
since $x$ again is no longer small when $y>6$, even at LHC energies.)

However, what we find is very different from the forward rapidity result\cite%
{BRAHMS} from RHIC for the final state hadrons where the average transverse
momentum per hadron is almost constant as rapidity increases. There might be
two reasons to explain this discrepancy. First, at RHIC the typical Bjorken $%
x$ is less than $10^{-2}$, and becomes comparable to 1 in the forward
rapidity region. Also, the saturation momentum for one of the colliding
nuclei becomes very small. This makes the saturation picture not applicable
to the RHIC data in that region. The second reason is that there exist
several intermediate processes (e.g., thermalization\cite{Bai01} and
hadronization) between the initial produced gluons and final measured
particles. This may indicate that the direct use of the parton-hadron
duality hypothesis might be questionable and dangerous in this case. In
order to get the right picture from initially produced partons to the final
state particles, we need to consider the intermediate processes like
thermalization which distorts the transverse momentum distribution of
produced partons.

Finally, we sketch how to deal with the case where the projectile and target
have different number of nucleons(e.g. D-Au collision). As one can see, $%
\frac{dN}{dy}$ reaches its maximum when two saturation momentum are the
same. This happens at $y=0$ in the $A_{1}=A_{2}$ case. While in the D-Au
collision, the critical rapidity can be obtained by setting
\begin{eqnarray}
Q_{s}(A_{1},-y+\frac{Y}{2})^{2} &=&Q_{0}(A_{1})^{2}\left( \frac{\sqrt{s_{1}}%
}{\sqrt{s_{0}}}\right) ^{\lambda _{s}}\exp \left[ -\lambda _{s}y\right]
\nonumber \\
&=&Q_{s}(A_{2},y+\frac{Y}{2})^{2}=Q_{0}(A_{2})^{2}\left( \frac{\sqrt{s_{2}}}{%
\sqrt{s_{0}}}\right) ^{\lambda _{s}}\exp \left[ \lambda _{s}y\right]
\end{eqnarray}%
and noting that the saturation momentum is proportional to $A^{\frac{1}{3}}$.

\section{Conclusion}

In the following, we would like to qualitatively discuss a little bit about
the consequences of the above results in forward rapidity region. As we have
seen(presuming $\delta$ is small compared to 1), the number of the produced
gluon and the transverse energy in per unit rapidity decrease considerably
in forward rapidity region, while the transverse energy per gluon per unit
rapidity increases a lot in forward rapidity region. This leads to some
changes compared to the central rapidity region.

1. Jet quenching: Jet quenching is a very important signal for the existence
of the quark gluon plasma (QGP) in relativistic heavy ion collisions. After
the hard scattering between two partons, each of the two scattered partons
evolves into a jet which consists of a leading particle together with a
number of associate hadrons around the leading particle during
hadronization. However, in the presence of dense strongly interacting matter
(QGP) surrounding the hard scattering site, the scattered partons have
induced gluon radiation due to multiple scattering with the plasma and this
causes the jet to lose energy. This energy loss causes a suppression of the
observed leading hadron spectrum. This is the jet quenching. However, since
the number of the produced gluons is much less in the forward rapidity
region than in the central rapidity region, the induced energy loss is
smaller since bremsstrahlung gluon radiation and multiple scattering are
expected to be less important. Therefore, we might expect that the jet
quenching effect would be less significant in the forward rapidity region.
However, in the very large rapidity region our discussion breaks down
because of two reasons: first, the saturation is no longer applicable to the
region where $x$ is large; the other reason is that most of the valence
quarks are in this region after the collision. The valence quarks may give a
significant contribution to partonic density and increase the induced strong
interaction which then might result in more energy loss.

2. Gluon thermalization: Baier $\mathit{et al.}$\cite{Bai01} have discussed
thermalization of the partons, after they are produced and before
hadronization, in the bottom-up thermalization scenario. In the above
picture, thermalization time can be expected to be considerably longer in
the regions away from the central rapidity as compared to the thermalization
at $y=0$.

3. The elliptic flow: The concept of the elliptic flow parameter $v_{2}$\cite%
{Ollitrault:1992bk,Huovinen:2001cy,Teaney:2001av,Gyu01} is introduced to
reflect the anisotropy of the produced gluons that due to the spatial
geometrical azimuthal asymmetry in the non-central collision. With the
similar picture and discussion for the jet quenching, we can see that the
elliptic flow is usually generated at the partonic level prior to the
hadronization. At RHIC, data exhibits a very strong elliptic flow and
saturate the hydrodynamic limit. This suggests that very rapid and almost
complete local kinetic equilibrium is reached. At LHC, elliptic flow is
again expected to saturate the hydrodynamic limit. In the forward rapidity
region at LHC energies due to the decrease of produced gluons, we may expect
a smaller elliptic flow.

In the saturation picture, we apply the McLerran-Venugopalan model
and the geometrical scaling to the computation of the multiplicity
and transverse energy distribution for heavy-ion collisions at LHC.
Results for the multiplicity and transverse energy distribution in
the central rapidity region agree well with the RHIC experimental
data and previous studies. Then we make predictions for the results
of the produced gluons at the early stage of heavy ion collisions in
both central and forward rapidity regions at the LHC energy scale.
Noticing that hadronization and thermalization can distort both the
multiplicity and transverse energy distribution, we need to consider
these effects before we compare the results of the produced gluons
with the data of the final state particles.

\textbf{Acknowledgements} I would like to thank my advisor Professor
Alfred.H. Mueller for suggesting this work and numerous stimulating
discussions. Without his patient guidance, this work would not be
possible. I also want to thank Cyrille Marquet for reading this
paper and providing helpful comments on this work.

\end{document}